# Enhancing Data Security in the User Layer of Mobile Cloud Computing Environment: A Novel Approach

Noah Oghenfego Ogwara[1], Krassie Petrova[2], Mee Loong (Bobby) Yang[3], Stephen G. MacDonell[4]

*School of Engineering, Computer and Mathematical Sciences*
*Auckland University of Technology, Auckland, New Zealand*

[1]fego.ogwara@aut.ac.nz, [2]krassie.petrova@aut.ac.nz, [3]bobby.yang@aut.ac.nz, [4]stephen.macdonell@aut.ac.nz

**Abstract**

*This paper reviews existing Intrusion Detection Systems (IDS) that target the Mobile Cloud Computing (MCC), Cloud Computing (CC), and Mobile Device (MD) environment. The review identifies the drawbacks in existing solutions and proposes a novel approach towards enhancing the security of the User Layer (UL) in the MCC environment. The approach named MINDPRES (Mobile-Cloud Intrusion Detection and Prevention System) combines a host-based IDS and network-based IDS using Machine Learning (ML) techniques. It applies dynamic analysis of both device resources and network traffic in order to detect malicious activities at the UL in the MCC environment. Preliminary investigations show that our approach will enhance the security of the UL in the MCC environment. Our future work will include the development and the evaluation of the proposed model across the various mobile platforms in the MCC environment.*

## 1. INTRODUCTION

Mobile Cloud Computing (MCC) is a distributed computing technology that improves on the resource limitations posed by the Mobile Device (MD). This technology leverages the benefits of the powerful computing resources offered by Cloud Computing (CC) Infrastructure [1]. MCC integrates CC services into the Mobile Computing (MC) environment; by outsourcing data to the cloud and allows MD to utilize less memory, but the security of user data during outsourcing may be questionable [2]. Extensive research and development effort has been carried out, with the aim to provide a more secure MCC environment and attract more consumers to use these services. However, security and privacy have been reported as a major challenge that hinders the MCC technology [3]- [5, [7-15].

Data security issues cut across the different models that make up the MCC architecture, including the User Layer (UL), the Communication Layer (CL), and the Cloud Service Provider Layer (CSPL) [7]. The sensitivity of the information stored or accessible at the UL makes it vulnerable to cybercrime attacks. Studies have reported that most MD users do not install protective applications such as anti-virus, anti-malware, anti-spyware, amongst others making them more vulnerable in the MCC environment [12-14]. Google's simple publishing policy about official Android applications (apps) makes their users a prime cybercrime target. Malicious apps have been uploaded successfully to the Google play store despite the protection of its inbuilt security model [15]. Furthermore, existing defensive solutions such as anti-virus for MDs are not efficient at eliminating the ever-increasing mobile malware threat. This is as a result of their total dependence on signature-based approach for detection. Furthermore, MDs are not suited for constant malware scanning due to their known resource limitations [14].

This paper presents the current state-of the art in IDS solutions in the cloud and mobile environment and proposes a model that provides a protection to the highly exposed UL. The paper builds on our earlier review which explored the MCC security solution landscape environment and suggested that Intrusion Detection Systems (IDS) provided a feasible approach towards addressing a wide spectrum of MCC security threats [76]. The rest of the paper is organized as follows. First, it provides a brief discussion about the nature of intrusion types in the MCC environment, followed by a comprehensive review of existing IDS solutions. Next, the paper identifies security issues that need to be addressed and proposes a model that may be used to build a security enhancing solution for the UL of MCC. Directions for further research and development are also highlighted.

## 2. INTRUSION AND INTRUSION DETECTION SYSTEMS

IDS monitor the activities that take place within the system or network in order to identify intrusions activities that violate the security policy of the systems. In MCC, intrusion can occur both at the MD, and in the cloud infrastructure.

### A. Intrusion in Mobile Devices

In the UL of the MCC infrastructure, MDs are used to access the resources of this environment. As the MD architecture is similar to the architecture of personal computers, MDs are vulnerable to the same types of intrusions and malicious activities, for example, viruses, trojan horses, spyware [16].

### B. Intrusion in the Cloud Infrastructure

MCC environments are also vulnerable to intrusion that targets the security of technology itself. confidential information that may be stored by users in cloud resources may become the target of an attack. Through obtaining unauthorized access attackers may violate the privacy and confidentiality of the data stored in the cloud by cloud users. Amongst others, the type of intrusion attacks that are prevalent in the CC infrastructure include insider attacks, flooding attacks, Denial of Services (DoS) attacks, user to root attacks, port scanning, Virtual Machine (VM)attacks, covert-channel attacks. Such intrusion attacks are dangerous since they affect both the MD users and the cloud service provider (CSP). Moreover, it is the responsibility of CSPs to provide adequate security protection of user information [16- [18].

### C. Types of IDS

There are mainly four different types of IDS used in Cloud: Host-based Intrusion Detection Systems (HIDS**),** Network-based Intrusion detection Systems (NIDS), Hypervisor based IDS (Hy-IDS), and Distributed Intrusion Detection Systems (DIDS).

**HIDS.** These IDS are installed in the host machine and detects intrusion by analyzing the information it receives. Information sources include for example, the file system, system calls, and log files.

**NIDS.** These IDS detect intrusion by analyzing the network packets in order to discern malicious activities in the network. NIDS compare the current network behaviour with previously observed behaviour to identify suspicious activities.

**Hy-IDS.** These IDS allow users to monitor the communication channels across VMs and analyze communication patterns in order to detect possible intrusion.

**DIDS.** These IDS may comprise several HIDS and NIDS over a large network. As DIDS deploy the detection techniques used by both NIDS and HIDS, DIDS inherit the benefits of both types of systems [16]-[17].

### D. IDS Detection Method

Detection Methods (DM) used in IDS can be categorized according to the following approach. Signature-based detection (SB) attempts to define a set of rules ('signatures'), which are used to detect (and predict consequentially) the appearance of known intrusion attack patterns. In a cloud-based system, this method can be used to identify a known attack [19].

Anomaly-based detection (AB) is concerned with identifying and labeling 'malicious' an event that deviate from the normal cloud network behavior. The method has the advantage of identifying attacks that may have not been found previously [17]. Finally, Hybrid detection (HB) is used to enhance the capabilities of an existing IDS by combining SB and AB in order to enable the detection of both known attacks, and unknown attacks [17].

## 3. METHODOLOGY AND RESULTS

A total number of 371 papers were obtained from our search across four different electronic databases (IEEE, Science Direct, ACM and Springer Link), published between 2010 and 2020. Only peer reviewed journal articles and conferences papers that were written in English language were used in this study. The 58 papers that were selected for a detailed review proposed either solutions or solution frameworks for intrusion and/or malware detection either in the MD, or in the CC/MCC environment. A summary of the review results is shown in Table 1. Each article is assigned an identifier and is referenced to the source paper (columns 1, 2, 3). The IDS type, its detection method and scope are shown in columns 4, 5, 6. The last three columns describe the characteristics, or dimensions, of the proposed solution or framework. A number of different criteria's have been used in prior research to investigate existing solutions and identify weakness [5-7],[17]. Extracted from the literature, the solution dimensions described below were used to analyze the IDS frameworks included in the review. The ✓ indicates that the framework includes the dimension in its architecture while ✘ indicates that the framework does not include it.

**IDS Type**. This dimension specifies the kind of IDS that was proposed in each framework or solution such as HIDS, NIDS, Hy-IDS, and DIDS.

**Detection Method**. This identifies the method of intrusion detection used in each proposed solution or framework, such as SB, AB and HB.

**Prevention Component**. This is used to indicate whether the framework or the solution provides a prevention technique whenever intrusion is detected.

**Machine Learning (ML) Component**. This specifies whether the framework incorporates any ML process or algorithm as part of the intrusion detection process.

**Performance Analysis**. This indicates whether the research discusses the performance and performance analysis methods associated with the proposed framework or solution.

### A. CC-based IDS Solutions and Frameworks

The solutions proposed in F2, F5, F47 and F49 are of the HIDS type, and target the CC environment. Frameworks F2 and F5 use an SB detection approach and proxy servers for in-depth forensic analysis of files stored locally at the device. The frameworks proposed in F47 and F49 use the AB detection approach with ML techniques. F47 uses a mobile agent to automatically collect data from each host for its detection process. However, F49 focuses on the protection of the VM in the CC environment. In this review, the NIDS proposed solution that use the SB detection method have their detection engine located at the cloud server. Only F21 and F36 contain attack prevention modules. The solution presented in F9 uses correlated alerts for its detection process. Some of the solutions that target the CC environments have shifted the detection target to the MD node as evidenced in F20 and in F30. The solution presented in F7 enhances the SB detection method by updating new signatures automatically. The NIDS frameworks that use the AB detection method for analyzing network traffic applies various ML techniques in detecting intrusions in the CC environment. F41 and F53 are concerned with detecting DoS attacks in the CC infrastructure. Framework F56 provides a novel technique for anomaly detection using the statistical features of time series. F50 applies both supervised and unsupervised ML to improve detection and classification of attacks in the CC environment. The security approach in F43 incorporates the uses of immune mobile agents while F44 uses correlation of alerts for a more effective detection of malicious activities in a network.

The solutions in F27, F40 and F42 also uses ML techniques. The solution in F42 is concerned with the prediction of the malicious device in the cloud network while F27 relies on the identification of a network path where disruption occurs as a result of longer transmission time and reduced speed in transmission, in order to detect intrusion.

The NIDS solutions presented in F37 and F18 address security issues concerning communication using a cloudlet controller, and a Virtual Private Network (VPN), respectively. However, the frameworks presented in F12, F6 and F16 analyze system calls and model the device behaviour to enable the IDS to identify attacks. The NIDS frameworks presented in F4, F10, F26, F46 and F51 apply the HB detection method to identify intrusions in the CC environment. Most of these solutions combine SNORT with ML algorithms. However, the solution in F46 uses honeypot technology to produce an early warning about possible threats and attacks.

Table 1. Results of the analysis of the selected IDS frameworks and solutions

| ID | Source | Year | IDS Type | Detection Method | Targeted Environment | Prevention Component | Machine Learning | Performance Analysis |
|---|---|---|---|---|---|---|---|---|
| F1 | [20] | 2011 | DIDS | AB | CC | ✗ | ✓ | ✗ |
| F2 | [21] | 2011 | HIDS | SB | CC | ✓ | ✗ | ✗ |
| F3 | [22] | 2011 | HIDS | AB | MD | ✗ | ✗ | ✗ |
| F4 | [23] | 2012 | NIDS | HB | CC | ✗ | ✓ | ✓ |
| F5 | [24] | 2012 | HIDS | SB | CC | ✓ | ✗ | ✗ |
| F6 | [25] | 2012 | NIDS | AB | CC | ✗ | ✓ | ✓ |
| F7 | [26] | 2012 | NIDS | SB | CC | ✗ | ✗ | ✗ |
| F8 | [27] | 2012 | DIDS | SB | CC | ✗ | ✗ | ✗ |
| F9 | [28] | 2012 | NIDS | SB | CC | ✗ | ✗ | ✗ |
| F10 | [29] | 2012 | NIDS | HB | CC | ✓ | ✓ | ✗ |
| F11 | [30] | 2013 | DIDS | HB | MD | ✓ | ✓ | ✗ |
| F12 | [31] | 2013 | NIDS | AB | CC | ✗ | ✗ | ✗ |
| F13 | [32] | 2014 | HIDS | AB | MD | ✓ | ✓ | ✓ |
| F14 | [33] | 2014 | HIDS | AB | MD | ✗ | ✓ | ✗ |
| F15 | [34] | 2014 | NIDS | AB | MD | ✗ | ✓ | ✓ |
| F16 | [35] | 2014 | NIDS | AB | CC | ✗ | ✓ | ✓ |
| F17 | [36] | 2014 | DIDS | HB | CC | ✗ | ✓ | ✗ |
| F18 | [37] | 2014 | NIDS | AB | CC | ✓ | ✓ | ✓ |
| F19 | [39] | 2014 | NIDS | AB | MD | ✗ | ✓ | ✓ |
| F20 | [40] | 2015 | NIDS | SB | CC | ✗ | ✗ | ✗ |
| F21 | [41] | 2015 | NIDS | SB | CC | ✓ | ✗ | ✗ |
| F22 | [11] | 2015 | Hy-IDS | AB | MCC | ✓ | ✗ | ✓ |
| F23 | [42] | 2015 | Hy-IDS | SB | CC | ✗ | ✗ | ✓ |
| F24 | [43] | 2015 | DIDS | AB | CC | ✗ | ✓ | ✗ |
| F25 | [44] | 2015 | Hy-IDS | AB | CC | ✗ | ✓ | ✗ |
| F26 | [45] | 2015 | NIDS | HB | CC | ✓ | ✓ | ✓ |
| F27 | [46] | 2016 | NIDS | AB | CC | ✗ | ✗ | ✗ |
| F28 | [47] | 2016 | HIDS | AB | MD | ✗ | ✓ | ✓ |
| F29 | [48] | 2016 | HIDS | AB | MD | ✗ | ✓ | ✓ |
| F30 | [49] | 2016 | NIDS | SB | CC | ✗ | ✗ | ✓ |
| F31 | [51] | 2016 | DIDS | HB | CC | ✓ | ✗ | ✗ |
| F32 | [52] | 2016 | Hy-IDS | AB | CC | ✗ | ✓ | ✓ |
| F33 | [53] | 2017 | DIDS | HB | CC | ✓ | ✗ | ✗ |
| F34 | [54] | 2017 | HIDS | HB | MD | ✗ | ✗ | ✓ |
| F35 | [55] | 2017 | Hy-IDS | AB | CC | ✗ | ✗ | ✗ |
| F36 | [56] | 2017 | NIDS | SB | CC | ✓ | ✗ | ✗ |
| F37 | [57] | 2019 | NIDS | AB | CC | ✓ | ✗ | ✓ |
| F38 | [58] | 2017 | DIDS | SB | CC | ✓ | ✗ | ✓ |
| F39 | [59] | 2017 | Hy-IDS | AB | CC | ✗ | ✗ | ✗ |
| F40 | [60] | 2017 | NIDS | AB | CC | ✗ | ✓ | ✓ |
| F41 | [61] | 2019 | NIDS | AB | CC | ✗ | ✓ | ✓ |
| F42 | [62] | 2018 | NIDS | AB | CC | ✓ | ✓ | ✓ |
| F43 | [63] | 2018 | NIDS | AB | CC | ✓ | ✗ | ✓ |
| F44 | [64] | 2018 | NIDS | AB | CC | ✗ | ✗ | ✓ |
| F45 | [38] | 2018 | NIDS | AB | MCC | ✗ | ✓ | ✓ |
| F46 | [65] | 2018 | NIDS | HB | CC | ✓ | ✗ | ✓ |
| F47 | [66] | 2018 | HIDS | AB | CC | ✗ | ✓ | ✓ |
| F48 | [67] | 2018 | DIDS | HB | CC | ✗ | ✗ | ✓ |
| F49 | [68] | 2018 | HIDS | AB | CC | ✗ | ✓ | ✓ |
| F50 | [69] | 2019 | NIDS | AB | CC | ✗ | ✓ | ✓ |
| F51 | [70] | 2018 | NIDS | HB | CC | ✗ | ✓ | ✓ |
| F52 | [50] | 2018 | NIDS | AB | MCC | ✗ | ✓ | ✓ |
| F53 | [71] | 2019 | NIDS | AB | CC | ✗ | ✓ | ✓ |
| F54 | [72] | 2019 | HIDS | AB | MD | ✗ | ✓ | ✓ |
| F55 | [6] | 2019 | DIDS | AB | MCC | ✗ | ✓ | ✓ |
| F56 | [73] | 2019 | NIDS | AB | CC | ✗ | ✗ | ✓ |
| F57 | [74] | 2020 | HIDS | AB | MD | ✗ | ✓ | ✓ |
| F58 | [75] | 2019 | HIDS | AB | MD | ✗ | ✓ | ✓ |

The solutions presented in F23, F25, F32, F35 and F39 are Hy-IDS. Only F23 uses the SB detection method while F25, F32, F35 and F39 use the AB detection in their detection engines, located at the CC infrastructure. However, none of the Hy-IDS use a hybrid detection method. The use of mobile agents is common to these solutions. These mobile agents carry intrusion alerts from each VM in the cloud to a management server for analysis, in order to detect distributed intrusion at the hypervisor layer. The solutions presented in F38, F8, F1, F24, F17, F31, F33 and F48 are DIDS. F8 and F38 use the SB detection method. The AB detection method was used in F1 and F24. The HB detection method was applied in F17, F31, F33 and F48.

**B. MD-based IDS Solutions and Frameworks**
The solutions proposed in F3, F13, F14, F28, F29, F34, F54, F57, and F58 are of the HIDS type and target the MD environment. In these frameworks, the detection engines are normally located at the device level except for F13 and F14. The solution presented in F13 focuses on device resource optimization and places its detection engine at the cloud. In contract, in F14, some parts of the detection engine are located at the device level while others reside at the cloud server. The HB detection method was applied in the solution presented in F34 while the other frameworks have adopted the AB detection method. However, none of the HIDS type that target the MD environment, apply the SB detection method. In F34, the dynamic and static analysis of malicious applications in the MD node using system calls is the adopted approach. In a similar fashion, F28 extracts system calls from the applications that reside on the devices, constructs a weighted directed graph, and applies a deep learning algorithm in order to detect new attacks. Framework F54 features an autonomous detection of malicious activities related to both known and unknown attacks, using ML techniques.

The frameworks presented in F3 and F13 use location-based services for detecting intrusion at the MD node. F14 runs a local malware detection algorithm at the MD node to check for a known malware family. The security solutions presented in F58 and F57 analyse system calls and system log files respectively, to determine if a given app is malicious or not. The security techniques in F29 revolve around using the Google Cloud Messaging service for malware detection. In this review, NIDS frameworks that target the MD environment were found in F15 and F19. Both frameworks use the AB detection method. The detection engine in F15 resides at the cloud while that of F19 resides at the device. The framework presented in F11 uses DIDS, with an HB detection approach. The detection engine in F11 resides at both the device and the cloud.

**C. MCC-based IDS Solutions and Frameworks**
The frameworks in F22, F45, F52, and F5 target the MCC environment. F22 and F55 are of the DIDS type while F45 and F52 are of the NIDS type. All the frameworks that target the MCC environment presented in this review apply an AB detection approach, using ML techniques. Only F22 has an attack prevention module. In the framework presented in F45, the attack detection module analyzes incoming requests and classifies each request as normal or suspicious based on a trained deep learning model. The security techniques used in F52 involve the application of principal component analysis and simulation in order to identify intrusion events in the MCC environment. The framework in F55 presents an ML-based IDS that secures data collection and data fusion in a distributed environment. The framework analyzes network traffic from different VM using a multi-layer intrusion detection approach. In the event of detected malicious activities, decisions are made about restricting access to a specific VM.

## 4. ISSUES IDENTIFIED IN THE REVIEW

The detailed analysis of the results of the review of the selected IDS solutions and frameworks shows that most of the existing IDS solution locate their detection engine at the cloud server. Most of the existing solutions gather and correlate alerts from different nodes, and forward network traffics to the cloud for analysis using a proxy server. Some require the duplication of real-life device into an emulator at the cloud server for an in-depth analysis. These common approaches, as seen in the existing solutions, may cause additional security issues.

First, the leakage of sensitive information during the forwarding of network traffic to a proxy server. Second, the relatively low effectiveness of the approach to centralize the correlation of alerts received from various nodes; this process may consume significant time before an attack is detected. In addition, this process also requires a constant connection to the server which might be disrupted at the device layer, for example due to lack mobile network coverage. Third, most of the existing solutions have adopted AB detection methods using ML techniques. This approach is associated with a high rate of false alarms. Nevertheless, the issue of insider attacks in the CC environment remains a serious issue; applying ML and deep learning techniques may help combat this threat. Furthermore, mitigation process was seen in few works, to manage intrusion events. Finally, the security issues in the MCC environment affecting the UL and the communication channel have not received much attention. However, attackers focus their attention on the UL which is relatively more exposed due to the lack of security awareness amongst its users. Attackers may target the UL using malicious applications, code obfuscation, and repackaging of popular and legitimate apps with a malicious payload that is difficult to detect by the existing defensive techniques. The model proposed below addresses the security issues identified, especially at the UL in the MCC environment. It adopts a hybrid IDS approach.

## 5. THE PROPOSED MODEL

The model proposed in this paper is called MINDPRES (Mobile-Cloud Intrusion Detection and Prevention System). It aims to enhance data security at the UL in the MCC environment using dynamic analysis of the device behavior and applying ML technique at run time (Figure 1). It is described below.

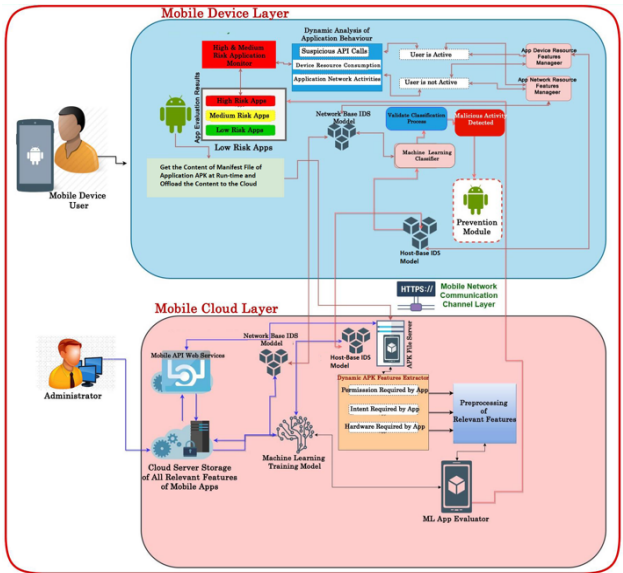

**Figure 1:** Mobile-Cloud Intrusion Detection and Prevention System (MINDPRES).

**A. Functional Description of the Model**

The model comprises three major components, namely: Application Evaluator, Detection Engine and Prevention Engine. The **application evaluator** is responsible for the preliminary assessment of each app that resides at the MD node, in order to ascertain the risk level of each user-installed application. The application evaluator extracts detailed information about the app from the manifest file of the application at run time, including the list of requested permissions, intent, and hardware's required. The extracted information is offloaded to the cloud where a risk assessment is conducted using an ML model.

The ML model would have been trained previously using preprocessed application data collected from different sources, with different ML classification algorithms. As a result of the risk assessment, each user installed app is classified as either of high or medium or low risk level; this result is sent as feedback to the device and the suspicious app watchlist is updated accordingly. The watchlist enables MINDPRES to monitor only the user-installed app included in the watchlist database rather than all user-installed apps. The **detection engine** uses a combination of an IDS and an Intrusion Prevention System (IPS) to safeguard the device. The HIDS dynamically monitors suspicious API calls to the device (e.g., device root access, installation of new application), and the device resource consumption (e.g., CPU usage, memory usage)) by the apps in the watchlist. The network activities of the apps in the watchlist are monitored by a NIDS, including Internet usage, requested URLs, upBytes, DownBytes, and other network activities. The device activities are monitored both when the device is active (i.e., used by its owner), and when the device is idle. The **prevention engine** in MINDPRES aims to provide an automatic mitigation process once an intrusion is detected. This will stop the execution of the detected application or block all malicious network traffic from a specific host. However, MINDPRES gives the user the flexibility to either allow the execution of the application if the user feels the application is safe for the device due to the possibility of false alarms.

**B. System Design Procedures**

In this phase, we intend to collect data from Android applications by downloading benign and malicious applications from different known source such as Google play store. We will submit each application to the VirusTotal engine to ascertain if the application is benign or malicious. The manifest file will be extracted from the apk of the application using the Android apk easy tool. Information from the manifest file of each application contains details of permissions requested, intent requested, and hardware resources required by each application. The collected data will be preprocessed, and a corresponding dataset will be generated. Thereafter, we will apply different data mining techniques to extract relevant features for distinguishing malicious activities in the MCC environment. instead of using all the features extracted. These relevant features will be stored in a central cloud server to build our ML model. The proposed model will be trained using different classifiers namely, Support Vector Machine, LightGbmClassifier, StochasticDualCoordinateAscentClassifier, Naive Bayes classifier and K-means classifier.

**C. Implementation and Evaluation Procedures**

The proposed model will be implemented using C#.net programming language with Xamarin Android in the Microsoft Visual Studio Integrated development environment. After the implementation of the propose model, the effectiveness of the proposed model will be verified with a testbed of over 1000 mobile applications that will be collected from different sources. The data collected will consists of both benign and malicious applications. These data will be pre-processed, and the data will be divided into two parts. Eighty percent of the data will be used for training and twenty percent of the data will be used for testing. The preprocessed data will be used to build an ML model for both the application evaluator and the detection engines. The evaluation procedures of the proposed model will be in two stages. First, a pre evaluation plan for the application evaluator using the confusion matrix. The classification accuracy and false alarm rate will be used as our validation metrics. The best ML classifier will be selected to build the final ML model. Second, the evaluation of the prototype system will involve real life experiment by installing the prototype system (MINDPRES) in the three devices. The apps that will be installed in each of the devices will be tested on VirusTotal to ascertain whether the app is benign or malicious. The performance overhead will be determine using a standard benchmark tool named quadrant standard edition app that is available at Google play store.

## 6. CONCLUSION AND FUTURE WORKS

In this paper, we highlight some security issues that MCC infrastructure are faced with. The issues were identified after a comprehensive review of literature of existing IDS solutions proposed in the CC, MD and MCC environments. One of the key issues identified is the vulnerabilities of the UL. Hence a novel approach for enhancing the security of the UL in the MCC architecture name MINDPRES was proposed. MINDPRES has an application evaluator that is trained with the ML classifiers. This evaluator ascertains the risk level of all user-installed applications that resides at the UL in the MCC environment. MINDPRES also contains a hybrid IDS that detect intrusion at the device level by constantly monitoring applications activities while the device is being used or idle and an IPS for managing intrusions detected. The proposed approach is the first of its kind in the MCC domain to the best of our knowledge. Only few works have been done using IDS and IPS in the MCC environment. Most of such work has not really focus on the UL. Also, no works in MCC has combines dynamic analysis of device behaviour at run time with user activities using ML techniques for protection of MCC resources against attacks. In our proposed model the detection engine resides locally at the device level. This eliminates the need for constant connection to a remote cloud for protection as proposed by most of the existing solution.

In the future, we intend to complete the development and implementation of the proposed model across different mobile platforms to test its effectiveness in solving security issues in the MCC environment.